\documentclass[doublecol]{epl2}
\usepackage{graphicx}
\usepackage{amsmath}
\usepackage{verbatim}
\usepackage{amssymb}
\usepackage{mathtools}
\usepackage{xspace}
\usepackage{float}
\usepackage{ifthen}

\newcommand{\pibeta}{\pit{\beta}}
\renewcommand{\pibeta}{\pi^\text{R}}
\newcommand{\pizero}{\pit{0}}
\renewcommand{\pizero}{\pi^{\text{T}}}
\newcommand{\sigmazero}{\sigma_{\text{T}}}
\newcommand{\sigmabeta}{\sigma_{\text{R}}}
\newcommand{\xzero}{x_0}
\newcommand{\xbeta}{x_\beta}

\newcommand{\PROCEDURE}[1]{\textbf{procedure}\ \sub{#1}}

\newcommand{\BRACE}[1]{
\;\;\;  \left\{\begin{array}{l}#1\end{array} \right.}
\newcommand{\IS}[2]{#1 \leftarrow #2}

\newcommand{\FOR}[1]{\textbf{for}\ #1 \textbf{: }}

\newcommand{\ENDPROCEDURE}{\text{------} \\ \vspace{-0.8cm}}

\newcommand{\IF}[1]{\textbf{if } #1 \textbf{: }}

\newcommand{\AND}{\text{and}}

\newcommand{\ELSE}{\textbf{else: }}

\newcommand{\OUTPUT}[1]{\textbf{output}\ #1}

\newcommand{\INPUT}[1]{\textbf{input}\ #1}
\newcommand{\COMMENT}[1]{\text{\footnotesize (#1)}}

\newcommand{\SET}[1]{\{#1\}}

\newcommand{\sub}[1]{\texttt{#1}}
\newcommand{\eq}[1]{eq.~(\ref{#1})}

\newcommand{\Eqq}[1]{Equation~(\ref{#1})}
\newcommand{\eqtwo}[2]{eqs.~(\ref{#1}) and~(\ref{#2})}
\newcommand{\eqfromto}[2]{eqs.~(\ref{#1}) to~(\ref{#2})}

\newcommand{\fig}[1]{Fig.~\ref{#1}}
\newcommand{\quot}[1]{``#1''}

\newcommand{\etcp}{\textrm{etc}}
\newcommand{\ie}{\textrm{i.e.}}

\newcommand{\rhomat}[4][]{\rho \lc #2,#3,#4\rc}
\newcommand{\NCAL}{\mathcal{N}}  
\newcommand{\expc}[1]{\exp \glc #1 \grc} 
\newcommand{\ran}{\texttt{ran}}
\newcommand{\ranb}[2][]{\ran_{#1} \! \glb #2 \grb}  
\newcommand{\lc}{\left(}  
\newcommand{\rc}{\right)}  

\newcommand{\glb}{\left(}  
\newcommand{\grb}{\right)}  
\newcommand{\glc}{\left[}  
\newcommand{\grc}{\right]}  

\newcommand{\TO}{,\ldots,}

\newcommand{\xtilde}{\tilde{x}}

\newcommand{\dd}[1]{\mathrm{d}{#1\ }}   

\newcommand{\Deltau}{\Delta_{\tau}}

\newcommand{\mean}[1]{\left\langle #1 \right\rangle}
\newcommand{\half}{\frac{1}{2}}

\newcommand{\prog}[1]{Alg.~\ref{alg:#1} (\sub{#1})}

\newcommand{\progg}[1]{Algorithm~\ref{alg:#1} (\sub{#1})}

\newcommand{\progn}[1]{Alg.~\ref{alg:#1}}

\newcommand{\pos}[2]{x^{\{#1 \}}_{#2}}
\newcommand{\postilde}[2]{\xtilde^{\{#1 \}}_{#2}}
\newcommand{\repo}{\texttt{PathConvergence}\xspace}
\newcommand{\pInf}{\ensuremath{p^\infty}\xspace}

\newcommand{\REF}[2][]{
        \ifthenelse{\equal {#1} {}}{Ref.~\cite{#2}}{Ref.~\cite[#1]{#2}}}

\newcommand\subfig[2]{{Fig.~\ref{#1}{#2}}}
\newcommand\subcap[1]{{(#1):}}

\title{
Path convergence in diffusion models
}

\author{Roi Holtzman\inst{1},
 Roman Beauvallet\inst{2}
\and Werner Krauth\inst{2,1}}
\shortauthor{R. Holtzman \etal}

\institute{                    
  \inst{1} Rudolf Peierls Centre for Theoretical Physics, Clarendon
    Laboratory, Oxford OX1 3PU, UK\\
  \inst{2} Laboratoire de Physique de l'École normale supérieure, ENS,
    Université PSL, CNRS, Sorbonne Université, Université Paris Cité,
    Paris, France
}

\abstract{We discuss diffusion-model paths interpolating between a target
distribution known only through $p$ patterns and a reference distribution that
can be sampled. These interpolating paths can be constructed symmetrically or
else in forward direction (often referred to as a \quot{noising}) from the
target patterns to the reference distribution or in backward direction (as a
\quot{denoising}) from the reference distribution to the patterns. For backward
paths with identical diffusion noise, we consider the path convergence in
number of patterns $p$ towards the path 
for infinitely many patterns. In a one-dimensional test case,
we show that this  convergence is on a scale $1/\sqrt{p}$, but with infinite
mean square deviation. We demonstrate that the path convergence allows for
extrapolation towards the $p=\infty$ path which samples the target distribution.
We provide a  proof-of-concept extrapolation algorithm and propose the
convergence and extrapolation of paths  as a possible strategy for density
estimation and generalization. We illustrate all our algorithms through
pseudo-codes and provide Python implementations.
}

\begin{document}
\maketitle

\newfloat{algorithm}{ht}{loa}
\floatname{algorithm}{Algorithm }

Sampling from a probability distribution comprises a number of fundamental tasks
in statistics. The sampling from an explicit probability distribution (for
example, the Boltzmann distribution of classical physics or the density matrix
in quantum physics), has given rise, among others, to the huge field of Monte
Carlo methods~\cite{Metropolis1953,LandauBinderBook2013,Levin2008,SMAC}. The
task of sampling from a probability distribution only known through $p$ samples
is known as the generalization problem. For low-dimensional continuous
distributions, the task is related to the field of density
estimation~\cite{Wasserman2004,Wasserman2006}: From the $p$ samples we may
estimate the probability distribution $\pi$ (that is, estimate its density), and
then sample the estimate of $\pi$ using Monte Carlo methods. In recent years,
diffusion models~\cite{sohl-dickstein2015, ho2020denoising, song2021scorebased,
song2019generative} have been successfully applied to the generalization problem
for high-dimensional \quot{target} distributions. Diffusion models connect the
$p$ patterns of the target distribution to a reference distribution, typically a
Gaussian, through an interpolating path referred to as a \quot{noising}. Then, a
sample from the reference distribution undergoes a backward process (referred to
as \quot{denoising}), usually implemented by a neural network, which outputs a
sample of the approximate target distribution.

In this Letter, we analyze diffusive interpolation paths $\SET{x_0 \TO \xbeta} $
between a target distribution $\pizero$ represented by $p$ patterns $x_0^\mu,
\mu = 1 \TO p$ and a reference distribution $\pibeta$ with samples $\xbeta \sim
\pibeta$, in the language of statistical mechanics~\cite{Feynman1972} and, in
particular, of path-integral Monte Carlo~\cite{Ceperley1995,SMAC}. We construct
statistically identical paths either symmetrically (without any direction) or
else in forward direction from one of the patterns $x_0^\mu$ into $\pibeta$
(\quot{noising}) or in backward direction from a sample $\xbeta\sim \pibeta$ to
one of the patterns $x_0^\mu$ (\quot{denoising}). For a one-dimensional target
distribution under identical realizations of the diffusion noise, we find that
the backward paths converge on a scale of $1 / \sqrt{p}$ to the infinite-$p$
(\pInf) path which connects samples of the two distributions. The mean square
deviation of finite-$p$ paths from the \pInf path is infinite, but we compute
the median deviation to high precision. We finally discuss the subject of
extrapolation, which builds on the established convergence of backward paths. We
explicitly show that, for identical realizations of diffusion noise, backward
paths for independent sets of $p$ and $q$ patterns from $\pizero$, together with
the backward path for the combined $p+q$ patterns yield an extrapolated path
that, on average, is closer to the \pInf path than the $p+q$ path itself. In the
conclusion, we discuss the possible use of the convergence and the extrapolation
of paths in the context of density estimation and generalization. This Letter
presents a number of proof-of-concept algorithms in the form of pseudo-codes.
Python implementations are provided in an associated open-source
repository~\cite{REPO}.

To set the stage, we consider the partition function, \ie\ the sum over all
probabilities, for the combined target and reference distributions:
\begin{align}
 Z &= \int \dd{x_0} \pizero(x_0) \int \dd{x_\beta} \pibeta(x_\beta).
\label{equ:ZInfinity}
\end{align}
As discussed, $\pizero$ is known only through the $p$ patterns, so that we
replace the first integral in \eq{equ:ZInfinity} by the indicator function of
patterns and couple the latter to $\pibeta$:
\begin{align}
Z &= \frac1p \sum_{\mu=1}^p \text{\textbf{1}}_\mu \int \dd{\xbeta}
\pibeta(\xbeta)
\label{equ:Zp}
\\
   &= \frac1p \sum_{\mu=1}^p \int \dd{\xbeta} \pibeta(\xbeta)
 \frac
 {\rhomat[free]{x_0^\mu}{x_\beta}{\beta}}
 {\rhomat[free]{x_0^\mu}{x_\beta}{\beta}}
 \label{equ:ZCoupling}
 \\
   &= \frac1p \sum_{\mu=1}^p
   \int \int \int \dd{x_1}
   \dd{x_2} \dots
    \dd{x_N} \pibeta(x_N) \times
   \notag
    \\
   &
   \!\!\!\!
   \!\!
   \times \!\!
 \frac
 {
 \rhomat[free]{x_0^\mu}{x_1}{\Deltau}
 \rhomat[free]{x_1}{x_2}{\Deltau} \dots
 \rhomat[free]{x_{N-1}}{x_N}{\Deltau}
 }
 {\rhomat[free]{x_0^\mu}{x_N}{\beta}}.
 \label{equ:PathIntegral}
\end{align}
Here, $N \Deltau = \beta$, and $x_0 \equiv x_0^\mu$ and $x_N \equiv \xbeta$. In
\eq{equ:ZCoupling}, we connect the two distributions by multiplying with and
dividing by what amounts to the imaginary-time density matrix  $\rho$ of a
quantum system. For a free quantum particle, it is
\begin{equation}
 \rhomat[free]{x}{x'}{\Deltau} = \frac{1}{\sqrt{2 \pi \Deltau}}
 \expc{-\half \frac{(x - x') ^2}{\Deltau}},
\label{equ:FreeFormula}
\end{equation}
in other words a Gaussian which connects the patterns $x^\mu$ to samples of the
reference distribution $\pibeta$. The denominator in \eq{equ:ZCoupling}, absent
in other formulations of diffusion models, guarantees that any sample $\xbeta$
of the reference distribution $\pibeta$ connects with equal weight to any
pattern $x_0^\mu$ and, moreover, guarantees the symmetry between $\pizero$ and
$\pibeta$ in \eq{equ:ZInfinity}. \Eqq{equ:PathIntegral} expands the density
matrix in the numerator of \eq{equ:ZCoupling} into a multiple integral by
repeated use of the convolution formula for general density matrices or,
equivalently, the Chapman--Kolmogorov equation,
\begin{equation}
 \rho(x, x', \tau) = \int \dd{x''}
\rhomat{x}{x''}{\tau'}
\rhomat{x''}{x'}{\tau - \tau'}.
\label{eq:ChapmanKolmogorov}
\end{equation}
The interpolation path $\SET{x_0 \TO \xbeta}$ in \eq{equ:PathIntegral} appears
in Feynman's path-integral formulation of quantum mechanics~\cite{Feynman1972}.
For the free particle in \eq{equ:FreeFormula}, the convolution formula of
\eq{eq:ChapmanKolmogorov} is
equivalent to the Gaussian bridge for $x''$ with end points $x$ and $x'$:
\begin{equation}
1 =
\int \dd{x''}
\frac
{ \rhomat{x}{x''}{\tau'}
\rhomat{x''}{x'}{\tau - \tau'}}
{\rho(x, x', \tau)}
\label{equ:GaussianBridge}
\end{equation}
from which we can sample $x''$ as a Gaussian with mean and standard deviation as
follows:
\begin{align}
 \mean{x''} & =
 x' (\tau - \tau ') / \tau  +  x'' \tau ' / \tau,  \\
 \sigma & = \sqrt{\tau' (\tau - \tau')/\tau}.
\end{align}
In this Letter, we restrict ourselves to the free-particle case and thus
encounter the Gaussian bridge repeatedly. Nevertheless, we keep in mind that
\eqfromto{equ:Zp}{equ:PathIntegral} apply to any interacting quantum system.

\begin{algorithm}
\newcommand{\algo}{symmetric-construction}
\begin{center}
$\begin{array}{ll}
&\PROCEDURE{\algo}\\
&\IS{x_0}{\sub{choice}\SET{x_0^1 \TO x_0^p}};\
\IS{x_\beta}{\sub{sample}(\pibeta)}\\
*&\IS{T}{\SET{0, \beta}}\\
&\FOR{\text{all neighboring $\tau^\pm \in T$}}\\
&\BRACE{
\IS{\tau}{(\tau^- + \tau^+)/2}\ \COMMENT{\quot{midpoint}}\\
\IS{T}{T \cup \SET{\tau}}\\
\IS{\mean{x_\tau}}{(x_{\tau^-} + x_{\tau^+})/2}\\
\IS{x_\tau}{\mean{x_\tau} + \sub{gauss}(\sqrt{\tau^+ - \tau^-})}\\
}\\
&\OUTPUT{\SET{x_0 \TO x_\beta}, T}\\
&\ENDPROCEDURE
\end{array}$
\end{center}
\caption{\sub{\algo}.
First generation of the hierarchical midpoint construction for the path in
\eq{equ:PathIntegral}. Subsequent generations re-inject the output after line
\quot{$*$} (see \REF{REPO} for a Python implementation).
}
\label{alg:\algo}
\end{algorithm}

In path-integral Monte Carlo~\cite{Ceperley1995,SMAC}, free paths $\SET{x_0 \TO
\xbeta}$ (\eqtwo{equ:PathIntegral}{equ:FreeFormula}) serve as proposal moves in
a Markov-chain context. They are often constructed hierarchically: One first
samples $x_0$ and $x_\beta$, then $x_{\beta/2}$, then $x_{\beta/4}$ and
$x_{3\beta/4}$ \etcp. The first generation, after sampling $\mu$ and $\xbeta$
independently, follows from the partition function in \eq{equ:ZCoupling}, with
fixed $x_0^\mu$ and $\xbeta$:
\begin{align}
Z|_{x_0^\mu, \xbeta} & = \frac{1}{p}
\frac
 {\rhomat[free]{x_0^\mu}{x_\beta}{\beta}}
 {\rhomat[free]{x_0^\mu}{x_\beta}{\beta}}\\
& \propto
\int \dd{x_{\frac{\beta}{2}}}
\underbrace{\frac
 {\rhomat[free]{x_0^\mu}{x_{\frac{\beta}{2}}}{\frac\beta2}
 \rhomat[free]{x_{\frac{\beta}{2}}}{x_\beta}{\frac\beta2}}
  {\rhomat[free]{x_0^\mu}{x_\beta}{\beta}}}_{
  \!\!\!\!
  \!\!\!\!
  \text{probability to sample
   $x_{\beta / 2}$ given $x_0^\mu, \xbeta$}
   \!\!\!\!
   \!\!\!\!
   } .
\label{equ:SymmetricFirst}
\end{align}
The Gaussian bridge in \eq{equ:SymmetricFirst} samples $x_{\beta/2}$, as
implemented in \prog{symmetric-construction}, and in the next generation
$x_{\beta/4}$ and $x_{3\beta/4}$, \etcp.

We can construct equivalent forward paths starting from a randomly sampled
pattern $x_0^\mu$. At a given stage, where $x_\tau$ and $x_0$ are already
constructed, we appropriately restrict the partition function of
\eq{equ:ZCoupling} and construct the step from $x_\tau$ to $x_{\tau + \Deltau}$
using \eq{eq:ChapmanKolmogorov}
\begin{align}
& \!\!\! Z|_{x_0^\mu, x_\tau}\! =\!
\int \dd{x_\beta} \! \! \pibeta(x_\beta)
\frac{
\rhomat{x_0^\mu}{x_\tau}{\tau}
\rhomat{x_\tau}{x_\beta}{\beta - \tau}
}
{
\rhomat{x_0^\mu}{x_\beta}{\beta}
}
\label{equ:Z-restricted-forward}
\\
&\propto
\int \dd{x_\beta}
  \underbrace{
  \pibeta(x_\beta)
  \frac{
  \rhomat{x_\tau}{x_\beta}{\beta - \tau}
  }
  {
\rhomat{x_0^\mu}{x_\beta}{\beta}
  }
  }_{\text{$\propto$ probability to choose $x_\beta$}}
  \int \dd{x_{\tau+\Deltau}}
  \times
  \label{equ:ForwardIntermediate}
  \\
\! \!  &
\underbrace{
\frac{
\rhomat{x_\tau}{x_{\tau+\Deltau}}{\Delta_\tau}
\rhomat{x_{\tau+\Deltau}}{x_{\beta}}{\beta - \tau -\Delta_\tau}
}
{
  \rhomat{x_\tau}{x_\beta}{\beta - \tau}
         }
         }_{\text{probability to sample $x_{\tau + \Deltau}$ given $x_\tau$ and
$x_\beta$}}
         \!.
\label{equ:ForwardFinal}
\end{align}
This results in $Z|_{x_0^\mu, x_\tau, x_{\tau + \Deltau}}$. The endpoint
$x_\beta$ of the Gaussian bridge from $x_\tau$ in \eq{equ:ForwardFinal} is
sampled again at the next iteration of the path construction.

In \eq{equ:ForwardIntermediate}, we may also inverse the order of integrations
and write it as $ \int\dd{x_{\tau+\Deltau}} \dd{x_\beta} \cdots$. What comes
after the $ \int\dd{x_{\tau+\Deltau}} \cdots$ is again the probability to sample
$x_{\tau + \Deltau}$. For a Gaussian $\pibeta$, this results in a single
Gaussian distribution for $x_{\tau + \Deltau}$, rather than in the integral over
Gaussians bridges with endpoints $\xbeta$ in
\eqtwo{equ:ForwardIntermediate}{equ:ForwardFinal}. With $\pibeta = \NCAL(0,
\sigmabeta)$, we find that $x_{\tau + \Deltau} \sim \NCAL(\mean{x_{\tau +
\Deltau}}, \sigma_{\tau} )$ with:
\begin{align}
\label{equ:forward-mean}
  \mean{x_{\tau + \Deltau}} &= x_\tau - \Deltau \frac{x_\tau(\beta -
\sigmabeta^2) + x_0^\mu \sigmabeta^2}{\beta(\beta - \tau) + \tau \sigmabeta^2}
\\
\label{equ:forward-std}
  \sigma_{\tau} &= \sqrt{\Deltau \glc 1 - \Deltau \frac{\beta -
\sigmabeta^2}{\beta(\beta - \tau) + \sigmabeta^2 \tau} \grc}.
\end{align}
Therefore we arrive at two implementations for the forward construction: The
one-step sampling implemented in \prog{forward-construction}, and a two-step
version (available in \REF{REPO}) which temporarily chooses $\xbeta$ as in
\eq{equ:ForwardIntermediate} and then uses a Gaussian bridge from $x_\tau$ to
$\xbeta$, for $x_{\tau+\Deltau}$ (see \eq{equ:ForwardFinal}). Both versions give
equivalent paths to \progn{symmetric-construction}. The forward construction is
a stochastic interpolation in the original sense of Lévy~\cite{Levy1940},
between patterns representing $\pizero$ and samples of $\pibeta$. We view the
entire path $\SET{x_0 \TO \xbeta}$ as an element of a high-dimensional sample
space, rather than a Markov chain in imaginary time $\tau$.

\begin{algorithm}
\newcommand{\algo}{forward-construction}
\begin{center}
$\begin{array}{ll}
\PROCEDURE{\algo}\\
\INPUT{\SET{\Deltau, x_0^\mu, x_\tau}}\\
\IS{\mean{x_{\tau+\Deltau}}}{\text{see \eq{equ:forward-mean}}}\\
\IS{\sigma_\tau}{\text{see \eq{equ:forward-std}}}\\
\IS{x_{\tau + \Deltau}}{\mean{x_{\tau + \Deltau}} + \sub{gauss}(\sigma_\tau)}\\
\OUTPUT{x_{\tau +\Deltau}}\\
\ENDPROCEDURE
\end{array}$
\end{center}
\caption{\sub{\algo}. Sampling of $x_{\tau + \Deltau}$ given $x_\tau$ for a
discrete path whose construction has started at a pattern $x_0^\mu$. The segment
$x_\tau \to x_{\tau + \Deltau}$ points towards a position $\xbeta$ that changes
with discrete (imaginary) time $\tau$.
}
\label{alg:\algo}
\end{algorithm}

\begin{figure}[htb]
\begin{center}
	\includegraphics[width=1\linewidth]{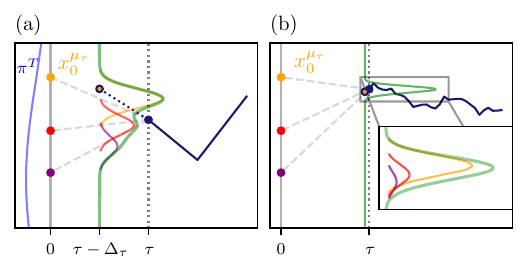}
\end{center}
\caption{Backward path construction.
\subcap{a} At finite $\Deltau$, the distribution of $x_{\tau - \Deltau}$
starting at $x_\tau$ is given by a sum of $p$ reweighted Gaussians
(in \emph{green}).
\progg{backward-construction} samples the pattern $\mu_{\tau}$ as the
endpoint $x_0^{\mu_\tau}$ of a Gaussian bridge for $x_{\tau - \Deltau}$.
\subcap{b} For small $\Deltau$, the $p$ Gaussians of (a) merge into a single
Gaussian (see \progn{backward-construction-dt}).
}
	\label{fig:PGaussiansOneGaussian}
\end{figure}
Interchanging $\pizero \Leftrightarrow \pibeta$ and $\tau \Leftrightarrow
\beta-\tau$ in the forward construction, we may construct backward  paths
starting from $x_\beta \sim \pibeta$. This again yields a restricted partition
function, which allows us to take a step to $x_{\tau - \Deltau}$:
\begin{align}
\label{equ:Z-restricted-backward}
\! Z|_{x_\tau, x_\beta} &=
\frac{1}{p} \sum_{\mu}
\frac{
\rhomat{x_0^{\mu}}{x_\tau}{\tau}
\rhomat{x_\tau}{x_\beta}{\beta - \tau}
}
{
\rhomat{x_0^{\mu}}{x_\beta}{\beta}
}\\
&\propto
\sum_{\mu_\tau}
\underbrace{
 \frac{
\rhomat{x_0^{\mu_\tau}}{x_{\tau }}{\tau }
  }{\rhomat{x_0^{\mu_\tau}}{x_\beta}{\beta}
  }}_{\text{$=\pi^{\mu_\tau}_\tau$ in \progn{backward-construction}}}
  \int \dd{x_{\tau-\Deltau}}
\label{equ:Z-restricted-backward-l1}
   \\
  &\times
\frac{
\rhomat{x_0^{\mu_\tau}}{x_{\tau - \Deltau}}{\tau - \Deltau}
\rhomat{x_{\tau -\Deltau}}{x_{\tau}}{\Delta_\tau}
}
{
\rhomat{x_0^{\mu_\tau}}{x_{\tau }}{\tau }
    }
    \!.
\label{equ:Z-restricted-backward-l2}
\end{align}
The integral over $\pibeta$  of Gaussian brigdes in \eq{equ:ForwardIntermediate}
becomes a weighted sum over patterns in \eq{equ:Z-restricted-backward-l1}.
\progg{backward-construction} evaluates the $p$ weights $\pi^{\mu_\tau}_\tau$, then samples the
temporary pattern $\mu_\tau$ from this resulting finite probability distribution
and finally samples $x_{\tau- \Deltau}$ from the Gaussian bridge of
\eq{equ:Z-restricted-backward-l2}. As indicated in the pseudo-code, we use two
random numbers for the construction of $x_{\tau - \Deltau}$, namely $\eta_\tau$
for the sampling of $\mu_\tau$, that is, the endpoint of the Gaussian bridge,
and $\gamma_\tau$ for the diffusion noise.

\begin{algorithm}
\newcommand{\algo}{backward-construction}
\begin{center}
$\begin{array}{cl}
&\PROCEDURE{\algo}\\
&\INPUT{\SET{\Deltau, x_\tau, x_\beta}, \SET{x_0^1 \TO x_0^p}}\\
&\FOR{\nu = 1 \TO p} \IS{\pi_\tau^\nu}{
\rhomat[free]{x_0^\nu}{x_\tau}{\tau}  /
\rhomat[free]{x_0^\nu}{x_\beta}{\beta} } \\
*&\IS{\mu_\tau}{\sub{sample}(\pi_\tau^1 \TO \pi_\tau^p)}\
\COMMENT{random number $\eta_\tau$}\\
&\IS{\sigma_\tau}{\sqrt{ \Deltau (\tau - \Deltau) / \tau}}\\
&\IS{\mean{x_{\tau-\Deltau}}}{(1 - \Deltau / \tau )  x_\tau +
\Deltau / \tau x_0^{\mu_\tau}
}\\
+&\IS{x_{\tau-\Deltau}}{\mean{x_{\tau-\Deltau}} + \sub{gauss}(\sigma_\tau)}\
\COMMENT{random number $\gamma_\tau$} \\
&\OUTPUT{x_{\tau - \Deltau}}\\
&\ENDPROCEDURE
\end{array}$
\end{center}
\caption{\sub{\algo}. Backward construction for arbitrary $\Deltau$ of a path
originating in $\xbeta$.
The Gaussian bridge for $x_{\tau - \Deltau}$ starts at $x_\tau$ and
ends at a pattern position $x^{\mu_\tau}_0$ that is sampled anew at each
$\tau$.}
\label{alg:\algo}
\end{algorithm}

The backward path constructs $x_{\tau - \Deltau}$ from $x_\tau$ and $\xbeta$
as a weighted sum over $p$ Gaussian bridges with endpoints $x_0^\mu$. The
distribution of $x_{\tau - \Deltau}$, the sum over $p$ Gaussians (see
\subfig{fig:PGaussiansOneGaussian}{a}) is thus best sampled in the two-step
procedure of \progn{backward-construction}. For small $\Deltau$, the means of
these Gaussians differ only on a scale $\sim \Deltau$, while their standard
deviations (which are all the same) are on a larger scale $\sim
\sqrt{\Deltau}$. In that limit, $x_{\tau - \Deltau}$ is Gaussian distributed
(see \subfig{fig:PGaussiansOneGaussian}{b}). It is sampled by a Gaussian
bridge from $x_\tau$ to the endpoint which is a weighted average of
the patterns, rather than by a single pattern $x_0^{\mu_\tau}$:
\begin{equation}
\underbrace{x_0^{\mu_\tau}}_{\text{in \progn{backward-construction}}}
\rightarrow
\mean{x_0^{\nu_\tau}} =
\underbrace{
 \glc  \sum_\mu
\pi_\tau^\mu
x_0^\mu \grc /
\sum_\mu \pi_\tau^\mu}_{\text{in \progn{backward-construction-dt}}}.
\end{equation}

\begin{figure}[htb]
\begin{center}
\includegraphics[width=1\linewidth]{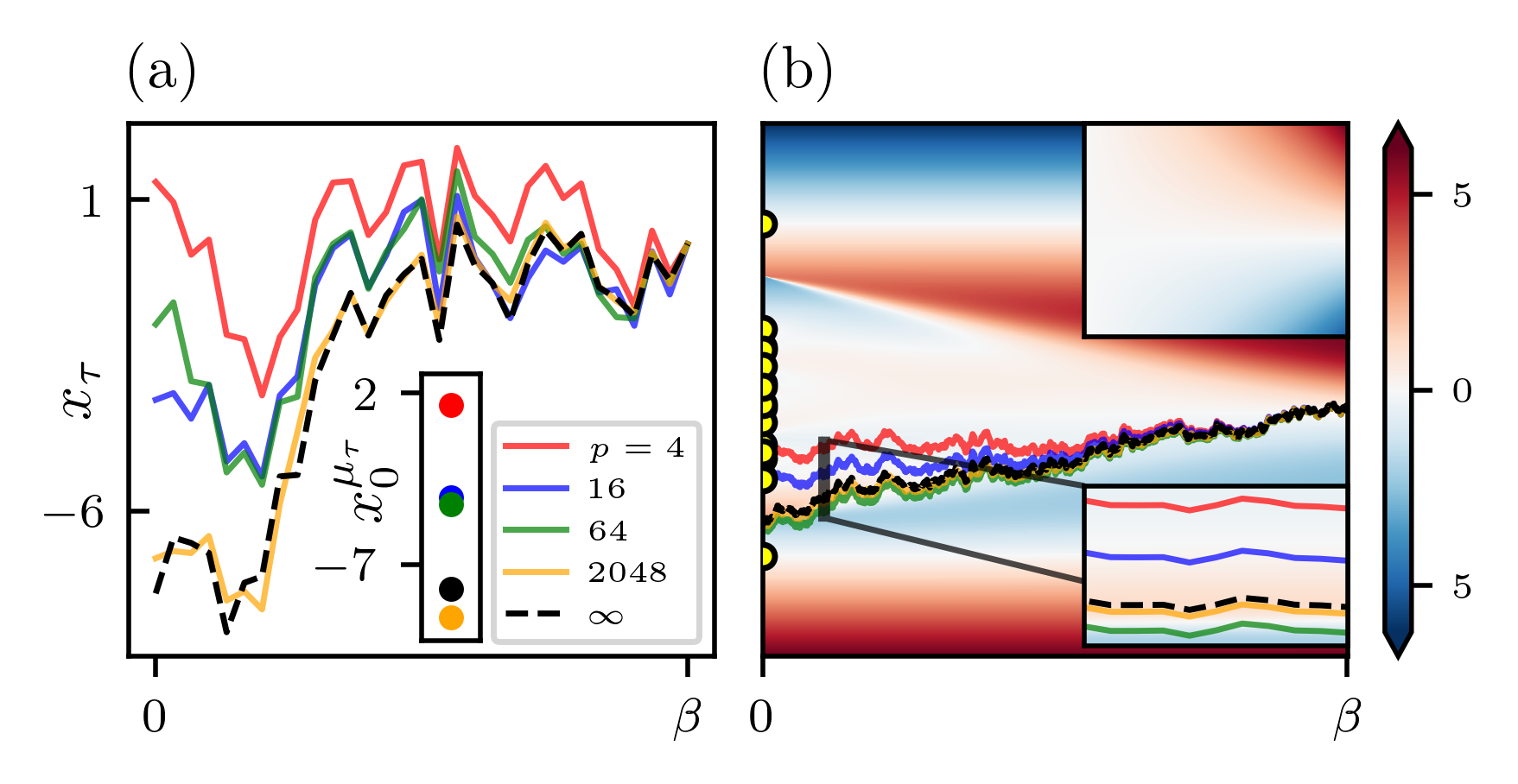}
\end{center}
\caption{Path convergence, velocity fields in our test case for $\beta=30$.
\subcap{a} Discrete backward paths for different $p$ under identical random
elements $(\eta_\tau,\gamma_\tau)$, from \prog{backward-construction}.
Endpoints $x_0^{\mu_\tau}$ are indicated for $\tau=6$.
\subcap{b} Continuous backward paths ($\Deltau=0.01$)
from \progn{backward-construction-dt}, with
velocity field
for $p=16$. Upper inset: velocity field for $p=\infty$. Lower inset: Closeup.
}
\label{fig:FlowPath}
\end{figure}

The backward construction of $x_{\tau - \Deltau}$ from $x_\tau$ in
\progn{backward-construction-dt} by the Gaussian bridge results in a path that
itself has no direction. However, the construction is equivalent to the
discretized backward motion $\Delta(x_{\tau - \Deltau}) = v^{\{p\}}_\tau(x_\tau)
\Deltau + \gamma_\tau$, where
\begin{equation}
v_\tau^{\{p\}}(x_\tau) = \tau^{-1}
\glb \frac{\sum_\nu \pi_\tau^\nu x_0^\nu}{\sum_\nu \pi_\tau^\nu} -x_\tau
\grb
\label{equ:VelocityField}
\end{equation}
is a velocity field and the random number $\gamma_\tau \sim
\sub{gauss}(\sigma_\tau)$  is a
Gaussian noise (see line \quot{$+$} in \prog{backward-construction-dt}). The
velocity field is analogous to the \quot{score} of diffusion models
\cite{hyvarinen2005estimation,vincent2011connection}, which however usually do
not incorporate the denominator in \eq{equ:ZCoupling}. The velocity field
$v_\tau^{\{p\}}(x_\tau)$ for a given set of $p$ patterns differs markedly from
the \pInf velocity field, as it has to guide the dynamics into one of the
patterns (see \subfig{fig:FlowPath}{b}, and inset). Diffusion models usually
attempt to learn $v_\tau^{\{\infty\}}(x_\tau)$ from $v_\tau^{\{p\}}(x_\tau)$
using a neural network.

\begin{algorithm}
\newcommand{\algo}{backward-construction-dt}
\begin{center}
$\begin{array}{cl}
&\PROCEDURE{\algo}\\
&\INPUT{\SET{x_\tau, x_\beta}, \SET{x_0^1 \TO x_0^p}}\\
*\!&\FOR{\nu = 1 \TO p} \IS{\pi_\tau^\nu}{
\rhomat[free]{x_0^\nu}{x_\tau}{\tau}  /
\rhomat[free]{x_0^\nu}{x_\beta}{\beta} }\\
&\IS{\sigma_\tau}{\sqrt{ \Deltau (\tau - \Deltau) / \tau}}\\
+\!&\IS{\mean{x_{\tau - \Deltau}}}
{ (1 - \Deltau /\tau) x_\tau\! +\! \Deltau/\tau \! \glc  \sum_\nu
\pi_\tau^\nu
x_0^\nu \grc \! / \!
\sum_\nu \pi_\tau^\nu } \\
&\IS{x_{\tau - \Deltau}}{\mean{x_{\tau - \Deltau}} + \sub{gauss}(\sigma_\tau)}\\
&\OUTPUT{x_{\tau - \Deltau}} \\
&\ENDPROCEDURE
\end{array}$
\end{center}
\caption{\sub{\algo}. Backward construction for small $\Deltau$
of a path originating in $x_\beta$.
The Gaussian bridge for $x_{\tau- \Deltau}$ starts at $x_\tau$ and
ends at a weighted average of the patterns.
}
\label{alg:\algo}
\end{algorithm}

\begin{figure}[htb]
\begin{center}
\includegraphics[width=1\linewidth]{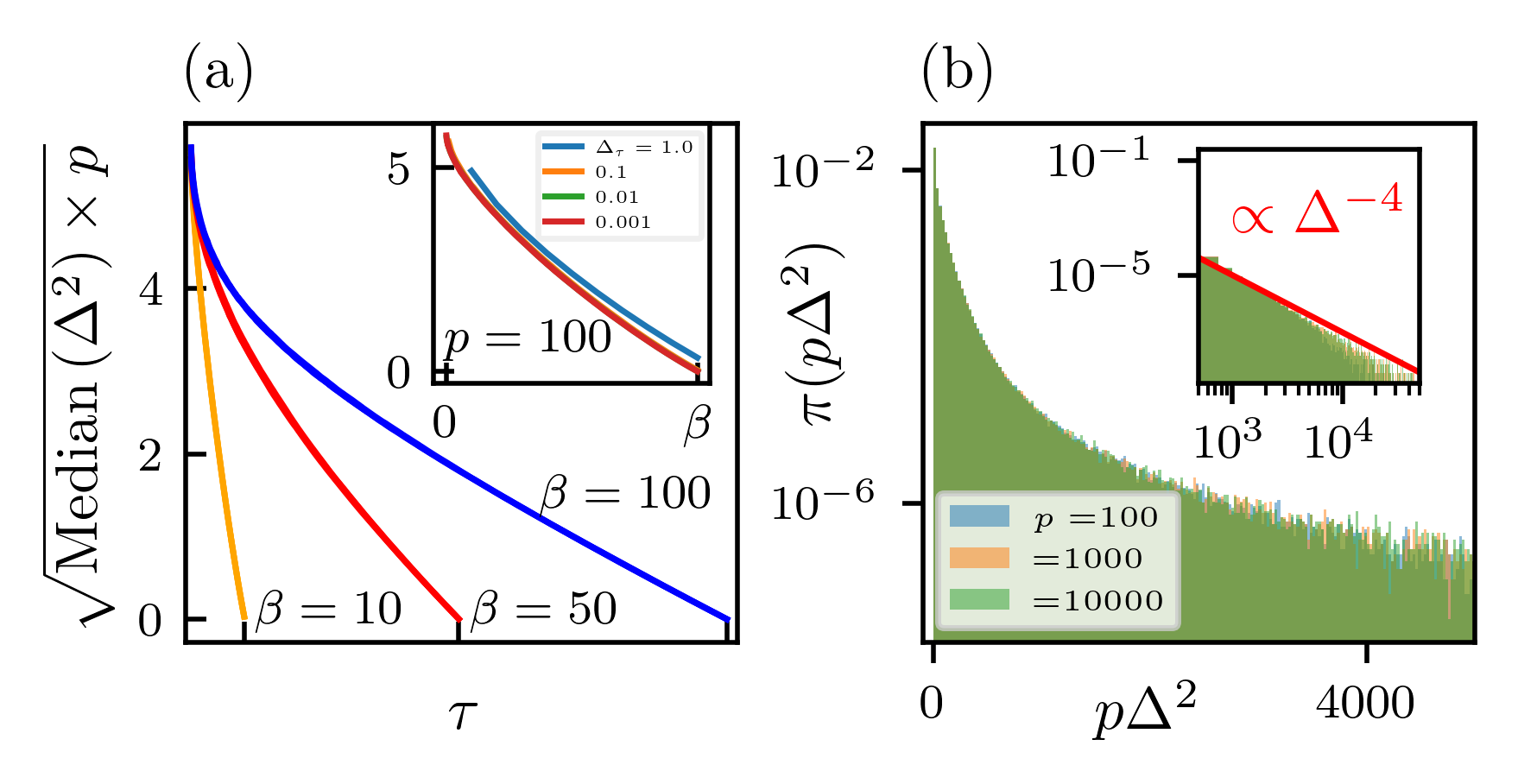}
\end{center}
\caption{Convergence of paths for our test case with
  \prog{backward-construction-dt}.
  (a) Median of $\sqrt{p \Delta^2}, \Delta = \pos{p}{\tau}- \pos{\infty}{\tau}$,
  for $\beta=10, 50, 100$ and $p=10, 100, 1000$ for $\Deltau=0.1$.
  Inset: same for $\beta=10, p=100$ and various $\Deltau$ illustrating
  convergence for $\Deltau \to 0$.
  (b) Distribution $\pi(\sqrt{p \Delta^2}) \sim \Delta^{-4}$ for various
  $p$.
  Inset: same at log scale.}
\label{fig:MedianUncorrected}
\end{figure}

In the backward construction of \progn{backward-construction}, corresponding to
finite-$\Deltau$, the position $x_{\tau - \Deltau}$ is constructed from $x_\tau$
using two random numbers. The first random number $\eta_\tau= \ranb{0, 1}$
samples the endpoint $x_0^{\mu_\tau}$ of the Gaussian bridge from $x_\tau$ with
probability $\pi^{\mu_\tau}_\tau$ in \eq{equ:Z-restricted-backward-l1} (see line
\quot{$*$} in the pseudo-code). The second random number $\gamma_\tau \sim
\mathcal{N}(0, \sigma_\tau)$ sets the diffusion noise (see line \quot{$+$} in
the pseudo-code). Using the same random number $\eta_\tau$ (that chooses the
pattern $x_0^{\mu_\tau}$) for different sets of patterns drawn from $\pizero$
can be done for one-dimensional patterns by ordering them as  $x_0^1 < x_0^2 <
\cdots < x_0^p$ and by constructing the cumulative distribution $\SET{\Pi_0=0,
\Pi_\nu = \Pi_{\nu-1} + \pi_\tau^\nu}$, and find the maximal $\mu$ such that
$\Pi_\mu < \eta_\tau \Pi_p $, that is, apply tower sampling~\cite{SMAC}.
Identical random numbers $\SET{\glb \eta_\tau, \gamma_\tau \grb}$ can be used
for different sets of patterns drawn from the distribution $\pizero$, as in the
coupling approach to Markov chains~\cite{ProppWilson1996,Levin2008}. For
increasing number $p$ of patterns, we notice the convergence of paths towards
the \pInf path (see \fig{fig:FlowPath}{a} for an illustration and
\REF{HoltzmanBeauvalletKrauth2} for an in-depth analysis of the finite-$\Deltau$
case).

In this Letter, we focus on the convergence of continuous paths of
\prog{backward-construction-dt},  with the diffusion noise, in other words the
Gaussians $\SET{\gamma_\tau}$,  as the only source of randomness. This case is
easier to generalize to higher-dimensional target distributions than the
construction of discrete paths (see \REF{HoltzmanBeauvalletKrauth2} for a
detailed discussion). The \pInf path that the finite-$p$ paths converge to is
constructed for any small $\Deltau$ by eliminating in
\progn{backward-construction-dt} the line \quot{$*$} and by replacing in line
\quot{$+$}:
\begin{equation}
\label{equ:p-to-pInf}
\glc \sum_\nu  \pi_\tau^\nu
x_0^\nu \grc /
\sum_\nu \pi_\tau^\nu
\to
\frac
{ \int \dd{\xzero}
\pizero \xzero
\frac{
\rhomat{\xzero}{x_\tau}{\tau}
}{
\rhomat{\xzero}{\xbeta}{\beta}}}
{ \int \dd{\xzero} \pizero
\frac{
\rhomat{\xzero}{x_\tau}{\tau}
}{
\rhomat{\xzero}{\xbeta}{\beta}}}.
\end{equation}
Here and in the following, we consider a test case  with $\pizero$ a
one-dimensional Gaussian of standard deviation  $\sigmazero =
5$ and zero mean, and $\xbeta=1$. In this test case,
convergence can be readily observed (see inset of \subfig{fig:FlowPath}{b}). For
this, we compare the paths $\pos{p}{\tau}$ to $\pos{\infty}{\tau}$ (for the same
small value of $\Deltau$) and compute the median of the squared deviation
$(\pos{p}{\tau} - \pos{\infty}{\tau})^2$. In our test case for $\pizero$, we
notice excellent precision of the median for a given value of $p$ and fixed
starting point $\xbeta$. Furthermore, the convergence with $\Deltau \to 0$ is
very good (see inset of \fig{fig:MedianUncorrected}{a}). Finally, the root
median square deviation (in other words, the median of the absolute deviation),
multiplied with $\sqrt{p}$ is virtually the same for all $p$ (see
\fig{fig:MedianUncorrected}{a}), demonstrating the convergence of the deviation
on its natural scale $\sim 1/\sqrt{p}$. Remarkably, we notice that the mean
squared deviation of paths appears to diverge. In fact, the probability
distribution of $\Delta^2 = (\pos{p}{\tau} - \pos{\infty}{\tau} )^2$, sampling
both the $p$ patterns $\{x_0^\mu\}$ and the diffusion noise $\{\gamma_\tau\}$,
scales as $\sim 1/\Delta^4$ (see \subfig{fig:MedianUncorrected}{b}), and has a
diverging mean on its scale $1/p$. The distribution of $p \Delta^2$ is largely
independent of $p$.

\begin{algorithm}
\newcommand{\algo}{backward-extrapolation-dt}
\begin{center}
$\begin{array}{ll}
&\PROCEDURE{\algo}\\
&\INPUT{
\SET{x_0^1 \TO x_0^p},
\SET{x_0^{1'} \TO x_0^{q}},
\SET{\gamma_{\Deltau} \TO \gamma_{\beta}}}\\
&\INPUT{
\pos{p}{\tau},
\pos{q}{\tau},
\pos{p+q}{\tau}
}\ \COMMENT{paths from $x_\beta$ with noise
$\mathbf{\gamma}$}\\
*&\IF{
\pos{p}{\tau}< \pos{p+q}{\tau}< \pos{q}{\tau} \AND\ \Delta_q \gg \Delta_p}\
\COMMENT{see \subfig{fig:SecondOrderExtrapolation}{a}} \\
&\BRACE{
\IS{\postilde{p+q}{\tau}}{\pos{p+q}{\tau} + \Upsilon}
}\ \COMMENT{$\Upsilon >0$}\\
&\ELSE\\
&\BRACE{
\IS{\postilde{p+q}{\tau}}{\pos{p+q}{\tau}}
}\\
&\OUTPUT{\postilde{p+q}{\tau}}\ \COMMENT{extrapolated path}\\
&\ENDPROCEDURE
\end{array}$
\end{center}
\caption{\sub{\algo}.
Extrapolation $\postilde{p+q}{\tau}$ from $\pos{p+q}{\tau}$ for paths
originating in $x_\beta$. See \eqtwo{equ:Ordering}{equ:CutoffSecondOrder} for
the precise conditions on $\Delta_p$ and $\Delta_q$.
}
\label{alg:\algo}
\end{algorithm}

The convergence of backward paths under identical diffusion noise is evidenced
by our comparisons of $\pos{p}{\tau}$ with increasing $p$ (see
\fig{fig:FlowPath}) and by the difference of  $\pos{p}{\tau}$ with
$\pos{\infty}{\tau}$ when the latter is available (see
\fig{fig:MedianUncorrected}). In generalization and density-estimation tasks,
the number $p$ of available patterns is severely limited.
In the remainder of this Letter, we consider two sets of $p$ independent
patterns from $\pizero$ that we refer to as $\SET{p}$ and $\SET{q}$,
respectively, with $\SET{p+q}$ used for the joint set of $2p$ patterns. For
concreteness, we again study our one-dimensional test problem, where we will
show that
the convergence of paths allows for their extrapolation.

\begin{figure}[htb]
\begin{center}
\includegraphics[width=\linewidth]{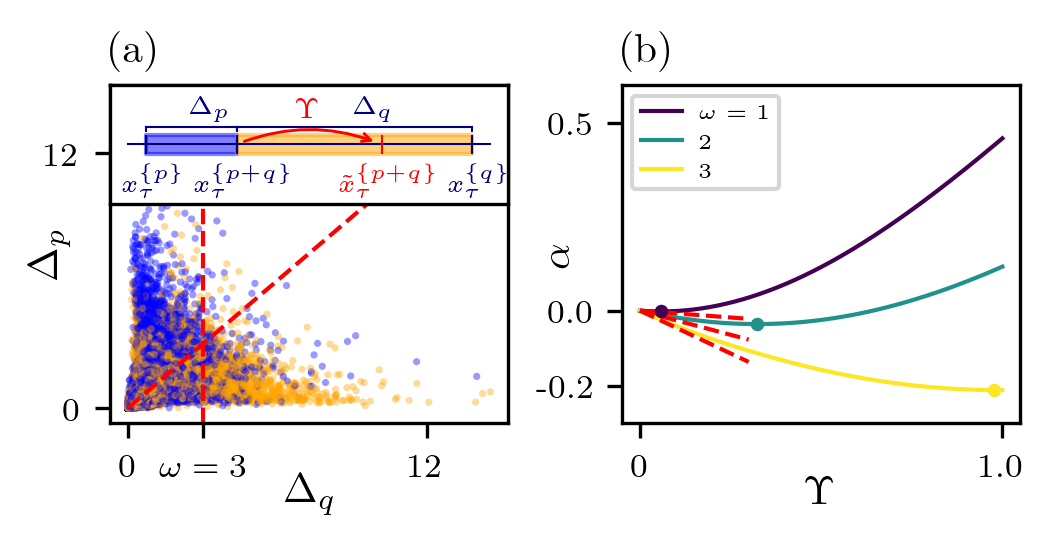}
\end{center}
\caption{Extrapolation for sets $\SET{p}$ and $\SET{q}$
of $p$ patterns joint into a set $\SET{p+q}$ (test case, $\tau=3$).
\subcap{a} Scatter plot for direction for $\pos{\infty}{\tau}$
with respect to $\pos{p+q}{\tau}$
(orange dots: $\pos{\infty}{\tau}$ towards $\pos{q}{\tau}$;
blue dots: $\pos{\infty}{\tau}$ towards $\pos{p}{\tau}$).
\subcap{b} Change $\alpha$ of difference with $\pos{\infty}{\tau}$ as a function
of $\Upsilon$ with linear approximation indicated (see
\eq{equ:ExtrapolationLinear}).
}
	\label{fig:SecondOrderExtrapolation}
\end{figure}

We consider paths $\SET{p}$, $\SET{q}$,  and $\SET{p+q}$ that satisfy, for a
given value of $\tau$:
\begin{equation}
\underbrace{\pos{p}{\tau} \,\,\,\,\,\, <
\,\,\,\,\,\,\,\,\,\,\,\,\,
}_{
\!\!\!\!\!\!\!\!\!\!\!\!\!
\Delta_p=
|\pos{p+q}{\tau} - \pos{p}{\tau}|
}
\!\!\!\!\!\!
\pos{p+q}{\tau}
\!\!\!\!\!\!\!
\!\!\!\!\!\!\!
\!\!
\!\!
\underbrace{
\,\,\,\,\,\,\,\,\,\,\,\,\,
 < \pos{q}{\tau}}_{
\,\,\,\,\,\,\,\,\,\,\,\,\,
\,\,\,\,
\Delta_q=
|\pos{q}{\tau} - \pos{p+q}{\tau}|
}.
\label{equ:Ordering}
\end{equation}
As there is an obvious relabelling $p \Leftrightarrow q$,
this leaves as an only
non-trivial condition that $\SET{p+q}$ is in between $\SET{p}$ and $\SET{q}$,
which
is satisfied in the majority of cases. It is natural that
for $\Delta_q > \Delta_p$, the \pInf path $\pos{\infty}{\tau}$
is more likely to be located between
$\pos{q}{\tau}$ and $\pos{p+q}{\tau}$ than between
$\pos{p+q}{\tau}$ and $\pos{p}{\tau}$.
We may tip the balance between
$\Delta_p$ and $\Delta_q$ defined in \eq{equ:Ordering} with a cutoff $\omega
>0$, and impose
the condition, among many other possible choices:
\begin{equation}
\Delta_q > \Delta_p; \quad \Delta_q > \omega.
\label{equ:CutoffSecondOrder}
\end{equation}
Then,
$\pos{\infty}{\tau}$ has higher probability
$p^+ (\omega) > \half$ to lie to the right of $\pos{p+q}{\tau}$ than to its
left. The probability $p^+(\omega)$ monotonically increases with $\omega$,
from $p^+(0) = \half$ all the way to $p^+(\infty) = 1$ (see
the scatter plot of  \subfig{fig:SecondOrderExtrapolation}{a}).

For small extrapolation parameters $\Upsilon>0$, we then set
\begin{equation}
\postilde{p+q}{\tau}(\Upsilon) =
\begin{cases}
\pos{p+q}{\tau} +\Upsilon& \text{if
\eq{equ:Ordering}, (\ref{equ:CutoffSecondOrder}) $\checkmark$}\\
\pos{p+q}{\tau} & \text{else}
\end{cases}.
\label{equ:PathExtrapolation}
\end{equation}
In the first case in this equation,
for $\Upsilon \gtrsim 0$, $\postilde{p+q}{\tau}$ improves the
representation of $\pos{\infty}{\tau}$
by an amount
$\Upsilon $ with probability $p^+(\omega)$ and deteriorates it by $\Upsilon$
with
probability $1 - p^+(\omega)$.
In the second case in this equation, the extrapolation has no effect and the
representation of $\pos{\infty}{\tau}$ neither improves nor deteriorates.
On average, under the conditions of \eqtwo{equ:Ordering}{equ:CutoffSecondOrder},
the distance to the \pInf path changes by
\begin{multline}
\overbrace{
\mean{
|\postilde{p+q}{\tau}(\Upsilon) -
\pos{\infty}{\tau}|}
-
|\pos{p+q}{\tau} - \pos{\infty}{\tau}  |
}
^{\alpha}
= \\
         - \glc 2 p^+(\omega) - 1 \grc \Upsilon\quad \text{for $\Upsilon \ll
1$},
\label{equ:ExtrapolationLinear}
\end{multline}
a relation that is easily checked numerically (see
\subfig{fig:SecondOrderExtrapolation}{b}). Here, negative $\alpha$ means that
the extrapolation is successful.

Our extrapolation algorithm is
readily implemented (see \prog{backward-extrapolation-dt}). The paths that are
effectively extrapolated as in the first case of \eq{equ:PathExtrapolation} are
on average closer
to $\pos{\infty}{\tau}$ with an improvement that is linear for small $\Upsilon$
(see \eq{equ:ExtrapolationLinear}) and that saturates when the extrapolation
parameter $\Upsilon$ tends to overshoot and that then diminishes. The algorithm
is
rudimentary, as it uses the restrictive conditions of
\eqtwo{equ:Ordering}{equ:CutoffSecondOrder} and compares paths
only at a fixed
value of $\tau$, \etcp. It is only meant to illustrate a point of principle,
namely that the convergence of random paths allows for their extrapolation. The
effect is clearly visible (see \subfig{fig:SecondOrderExtrapolation}{a}).

In conclusion, we have discussed interpolation paths
between probability distributions in a setup patterned after that
of modern diffusion models. The latter generically use neural networks to learn
score functions \cite{hyvarinen2005estimation, vincent2011connection,
song2021scorebased, ho2020denoising, albergo2025stochastic} aimed at constructing
positions (corresponding to our $\pos{p}{\tau}$ for $\tau \gtrsim 0$) that
approximately sample the target distribution $\pizero$ \cite{zhang2024minimax,
  chen2022sampling, oko2023diffusionminimax}.
In this Letter, in contrast, we have restricted our attention
to paths that exactly interpolate between the reference
distribution $\pibeta$ and the $p$ patterns $\{x_0^\mu\} \sim \pizero$. Discrete and
continuous paths were constructed without any approximation and without any of
the smoothings of patterns that are a staple in kernel-density
methods~\cite{Wasserman2006} yet problematic in high dimensions
\cite{biroli2026kernel, scarvelis2023closed}.
For the backward construction, our observed convergence in $1/\sqrt{p}$ under
identical diffusion noise is towards a path that reinstalls the symmetry between
$\pizero$ and $\pibeta$ and that samples both distributions and, in particular,
the target distribution $\pizero$.
The observed convergence, which might be of interest in its own right,
can likely be understood rigorously and much expanded on. We have
found, for
example, that the median absolute deviation for different values of $\beta$ is
a function of $\tau/\beta$ for $\tau/\beta \lesssim 1$ and that a rescaling of
patterns so that their empirical mean and variance match those
of $\pizero$ all but suppresses the deviation for that range of imaginary
times~\cite{HoltzmanBeauvalletKrauth2}. In both cases, the root mean square
deviation appears to be infinite on the scale $1/\sqrt{p}$. The convergence of
the continuous ($\Deltau \to 0$) paths seems robust in higher dimensions, and it
would be interesting to understand whether the same property can be preserved
for a discrete-time construction.

By definition, our exact backward paths do not generalize at fixed $p$. In fact,
from an arbitrary starting position $\xbeta$, they end up at any of the $p$
patterns with equal probability. Nevertheless, we have suggested that a given
set of $p +q$ patterns may be subdivided into smaller sets in an attempt to
extrapolate paths that all converge towards the same \pInf path. Our
proof-of-concept algorithm is naive: It considers a one-dimensional
target distribution, and only a single subdivision of
$p+q$, a restrictive conditioning (see line \quot{$*$} in
\progn{backward-extrapolation-dt}) at a fixed value of $\tau$. However, we
improve on the approximation of $\pos{\infty}{\tau}$ that is provided by
$\pos{p+q}{\tau}$. That this is at all possible is our point of principle,
although the effect is, for the moment, quite small.
Much work remains to be done to understand whether the
extrapolation of paths might be useful for generalization and
density-estimation tasks and whether a more complete setting of multiple
subdivisions of $p+q$ and of simultaneous extrapolations for all $\tau$
again requires neural networks to handle the vast amount of data
that might arise.

This Letter is accompanied by the \repo software package~\cite{REPO}
which is published as an open-source project under the GNU
GPLv3 license. \repo is available on GitHub as a part of
the JeLLyFysh organization. It contains Python implementations
of all the programs discussed here as well as a number of Mathematica notebooks.

\acknowledgments
We thank Philipp Hoellmer and  Giulio Biroli for helpful discussions and Shivaji
Sondhi for precious advice and continuous support. R.H. acknowledges support
from the Leverhulme Trust International Professorship Grant (No. LIP-2020-014).
R.B. was supported by ANR PRAIRIE-PSAI (France 2023) ANR-23-IACL-0008. W.K.
acknowledges generous support by the Leverhulme Trust.

\end{document}